\documentstyle[prl,eqsecnum,aps,floats,epsfig,amssymb]{revtex} % use this for PRL mode
\def\fsec    {\rm{fs}}                          % define how to write fsec
\def\MeV     {{\mbox{$\mathrm{MeV}$}}}
\def\GeV     {{\mbox{$\mathrm{GeV}$}}}

\def\Ntot    {162}                              % events in signal + sidebands
\def\Nevt    {22}                               % events in signal region
\def\Nside   {140}                              % events in sidebands 
\def\Nback   {6.1}                              % fit background events
\def\dBack   {0.5}                              % fit backround uncertainty
\def\Nsig    {15.9}                             % excess
\def\Signf   {6.3 \sigma}                       % Gaussian significance
\def\Mass    {3519}                             % Mass
\def\dMass   {1}                                % Mass uncertainty
\def\Width   {3}                                % Width
\def\dWidth  {1}                                % Width uncertainty
\def\Res     {\sim5}                            % Mass resolution
\def\lc      {{\mbox{$\Lambda_{c}^{+} $}}}
\def\lcpki   {{\mbox{$\Lambda_{c}^{+} \rightarrow pK^-\pi^+ \ $}}}
\def\kpi     {{\mbox{$K^- \pi^+$}}}
\def\pkpi    {{\mbox{$pK^- \pi^+$}}}

\def\ccd     {{\mbox{$\Xi_{cc}^{+}$}}}
\def\ccdlcki {{\mbox{$\ccd \rightarrow \lc K^-\pi^+ $}}}
\def\ccdlcko {{\mbox{$\ccd \rightarrow \lc \overline{K}^0 \pi^0 $}}}
\begin{document}

\title{
\rightline{FERMILAB-Pub-02/183-E}\vfill
First Observation of the Doubly Charmed Baryon \mbox{\boldmath $\ccd$}}

\author{
M.~Mattson$^{3}$,
G.~Alkhazov$^{11}$,
A.G.~Atamantchouk$^{11}$$^{,\ast}$,
M.Y.~Balatz$^{8}$$^{,\ast}$,
N.F.~Bondar$^{11}$,
P.S.~Cooper$^{5}$,
L.J.~Dauwe$^{17}$,
G.V.~Davidenko$^{8}$,
U.~Dersch$^{9}$$^{,\dag}$,
A.G.~Dolgolenko$^{8}$
% ...
\\                                                                            
\vskip 0.50cm                                                                 
\centerline{(SELEX Collaboration)}   
\vskip 0.50cm                                                                 
}

\date{August 9, 2002}
\address{
$^1$Ball State University, Muncie, IN 47306, U.S.A.\\
$^2$Bogazici University, Bebek 80815 Istanbul, Turkey\\
$^3$Carnegie-Mellon University, Pittsburgh, PA 15213, U.S.A.\\
$^4$Centro Brasileiro de Pesquisas F\'{\i}sicas, Rio de Janeiro, Brazil\\
$^5$Fermi National Accelerator Laboratory, Batavia, IL 60510, U.S.A.\\
$^6$Institute for High Energy Physics, Protvino, Russia\\
$^7$Institute of High Energy Physics, Beijing, P.R. China\\
$^8$Institute of Theoretical and Experimental Physics, Moscow, Russia\\
$^9$Max-Planck-Institut f\"ur Kernphysik, 69117 Heidelberg, Germany\\
$^{10}$Moscow State University, Moscow, Russia\\
$^{11}$Petersburg Nuclear Physics Institute, St. Petersburg, Russia\\
$^{12}$Tel Aviv University, 69978 Ramat Aviv, Israel\\
$^{13}$Universidad Aut\'onoma de San Luis Potos\'{\i}, San Luis Potos\'{\i}, Mexico\\
$^{14}$Universidade Federal da Para\'{\i}ba, Para\'{\i}ba, Brazil\\
$^{15}$University of Bristol, Bristol BS8~1TL, United Kingdom\\
$^{16}$University of Iowa, Iowa City, IA 52242, U.S.A.\\
$^{17}$University of Michigan-Flint, Flint, MI 48502, U.S.A.\\
$^{18}$University of Rome ``La Sapienza'' and INFN, Rome, Italy\\
$^{19}$University of S\~ao Paulo, S\~ao Paulo, Brazil\\
$^{20}$University of Trieste and INFN, Trieste, Italy\\
}
\maketitle

\begin{abstract}
  We observe a signal for the doubly charmed baryon $\ccd$
  in the charged decay mode $\ccdlcki$ in data from SELEX, the charm 
  hadro-production experiment at Fermilab.  We observe an excess of $\Nsig$ 
  events over an expected background of $\Nback \pm \dBack$ events, a 
  statistical significance of $\Signf$.  The observed mass of this state is 
  $\Mass \pm \dMass~\MeV /c^2$.  The Gaussian mass width of this state is 
  $\Width~\MeV /c^2$, consistent with resolution; its lifetime is less than 
  33~$\fsec$ at 90\% confidence.

\end{abstract}
\vfill

\twocolumn

% Introduction
%

The addition of the charmed quark to the ($uds$) triplet extends the 
flavor symmetry of the baryon octet and decuplet from SU(3) to
SU(4).  Even though the large ${m_c}$ breaks the symmetry, SU(4) still 
provides a good classification scheme for baryons composed of $uds$ and 
$c$~\cite{PDG}.  There is strong experimental evidence for all 
the predicted baryon states which contain zero or one valence charmed 
quark~\cite{PDG}.  In this letter we present the first experimental evidence 
for one of the six predicted baryon states which contain two valence charmed 
quarks - the doubly charmed baryons.  There have been many predictions of the 
masses and other properties of these states~\cite{HQEH,SW,Korner,BPhys}.  
The properties of doubly charmed baryons provide a new window into the 
structure of baryonic matter.

% SELEX description
%

The SELEX experiment uses the Fermilab 600 $\GeV /c$ charged hyperon beam  to
produce charm particles in a set of thin foil targets of Cu or diamond.  The
three-stage magnetic spectrometer is shown elsewhere~\cite{Thesis,SELEX}.  The
most important features are: (a) the high-precision, highly redundant, silicon 
vertex detector that provides an average proper time resolution of 20~$\fsec$ 
for single-charm particle decays, (b) a 10 m long Ring-Imaging Cherenkov 
(RICH) detector that separates $\pi$ from K up to 165 $\GeV /c$~\cite{RICH}, 
and (c) a high-resolution tracking system that has momentum resolution of 
${\sigma}_{P}/P<1\%$ for a \mbox{200\,$\GeV /c$} reconstructed 
$\Lambda_{c}^{+}$.

The experiment selected charm candidate events using an online
secondary vertex algorithm.  A scintillator trigger demanded an
inelastic collision with at least four charged tracks in the
interaction scintillators and at least two hits in a positive
particle hodoscope after the second analyzing magnet.  Event selection
in the online filter required full track reconstruction for measured
fast tracks ($p\,{\scriptstyle\gtrsim}$15 $\GeV /c$).  These
tracks were extrapolated back into the vertex silicon planes and
linked to silicon hits.  The beam track was measured in upstream
silicon detectors.  A full three-dimensional vertex fit was then
performed.  An event was written to tape if all the fast tracks in the
event were inconsistent with having come from a single primary
vertex.  This filter passed 1/8 of all interaction triggers and had
about $50\%$ efficiency for otherwise accepted charm decays.  The
experiment recorded data from $15.2 \times 10^{9}$ inelastic
interactions and wrote $1 \times 10^{9}$ events to tape using both
positive and negative beams. $67\%$ of events were induced by $\Sigma^{-}$,
$13\%$ by $\pi^{-}$, and $18\%$ by protons.

This analysis began with a sample of $\Lambda_c^+$ single-charm baryons
decaying to $\pkpi$.  Candidates were selected with a topological 
identification of 3-prong positively-charged secondary vertices, requiring a
momentum measurement for each track.  RICH identification of the proton and 
kaon was required. Charged tracks with reconstructed momenta which traversed 
the RICH ($p\,{\scriptstyle\gtrsim}\,22\,\GeV/c$) were identified as
protons or kaons if those hypotheses were more likely than the pion 
hypothesis.  The other positive track was identified by the RICH as a pion 
when possible, otherwise it was assumed to be a pion.  The primary vertex was 
refit using all other found tracks.  Details of the selection procedure can be 
found in~\cite{E781-PRL,E781-PL}. This $\lc$ sample (1630 events) contains the 
same $\lcpki$ events we used to measure the $\lc$ lifetime~\cite{E781-PRL}.

% Search Strategy
%

A Cabibbo-allowed decay of a doubly charmed baryon must have a 
net positive charge and contain a charmed quark, a strange quark and a baryon. 
We chose to search for decay modes like $\ccdlcki$ with an 
intermediate $\kpi$ secondary vertex between the primary vertex and the $\lc$ 
vertex.  In this analysis we have incorporated the charm selection techniques 
developed for single-charm baryon states, as above.

\begin{figure}[h]
%\ref{fig:event}
\center{\psfig{figure=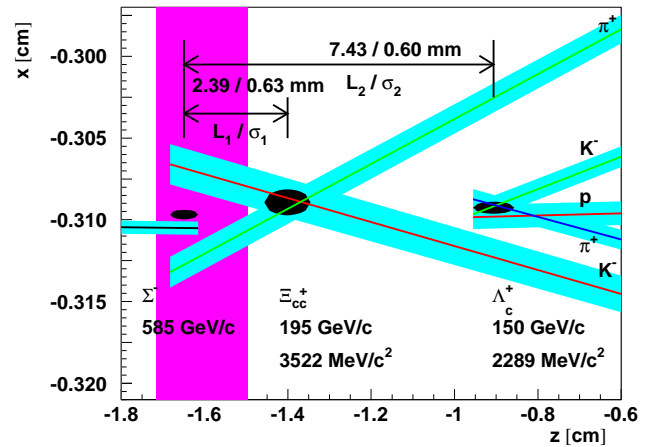,width=95mm}}
\caption{A candidate event with production target, $ 1\sigma$ track error 
         corridors and vertex error ellipses.  This is a plan view of three 
         dimensional tracks and vertices. Three additional found tracks which 
         form the primary vertex with the beam track are not shown.}
\label{fig:event}
\end{figure}

Events were analyzed for evidence of a secondary vertex composed of an
opposite-signed pair between the primary and the $\lc$ decay point.  
We used all tracks not assigned to the $\lc$ candidate in the search.  A new 
primary vertex was formed from the beam track and tracks assigned to neither 
the $\lc$ nor the $\kpi$ vertices.   The new secondary vertex had to have an 
acceptable fit $\chi^2$ and a separation of at least 1$\sigma$ from the new 
primary.  The $\lc\kpi$ transverse momentum with respect to the incident beam 
direction is required to be in the range $0.2 < p_t [\GeV/c] <2.0$.   These 
cuts were developed and fixed in previous searches for short-lived 
single-charm baryon states.  We have applied them here without change.  

Most tracks from the $\kpi$ vertex have insufficient momentum to reach 
the RICH.  Masses were assigned according to topology.  For the signal channel 
negative tracks are assigned the kaon mass and positive tracks the pion mass.  
As a background check we also kept wrong-sign combinations in which the mass 
assignments are reversed.  A candidate event from the $\lc K^- \pi^+ $
sample is shown in Fig.~\ref{fig:event}. Further details of the $\lc$ 
reanalysis may be found in Ref.~\cite{Thesis}.

% Search results and significance
%

\begin{figure}[ht]
%\ref{fig:mass}
\center{\psfig{figure=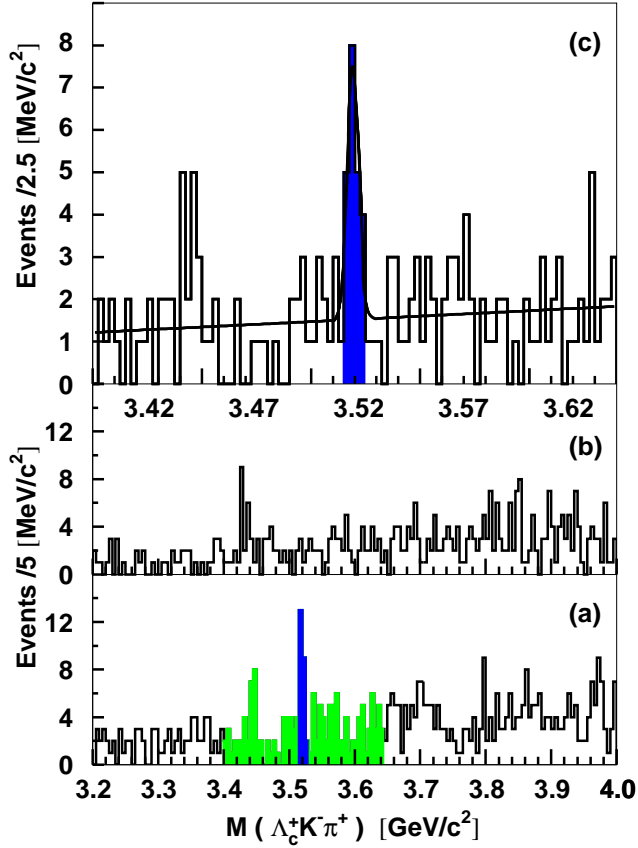,width=95mm}}
\caption{(a) The $\lc K^- \pi^+$ mass distribution in 5 $\MeV /c^2$ bins.
             The shaded region 3.400-3.640 $\GeV /c^2$ contains the signal 
             peak and is shown in more detail in (c).
         (b) The wrong-sign combination  $\lc K^+ \pi^-$ mass distribution in 
             5 $\MeV /c^2$ bins.
         (c) The signal (shaded) region ($\Nevt$ events) and sideband mass 
             regions with $\Ntot$ total events in 2.5 $\MeV /c^2$ bins.  
             The fit is a Gaussian plus linear background.}
\label{fig:mass}
\end{figure}

In Fig.~\ref{fig:mass}(a) we plot the invariant mass of the $\lc K^- \pi^+$
system, fixing the $\lc$ mass at 2284.9 $\MeV /c^2$~\cite{PDG}. The data, 
plotted in 5 $\MeV /c^2$ bins, show a large, narrow excess at 
3520 $\MeV /c^2$.  This excess is stable for different bin widths and bin 
centers.  Fig.~\ref{fig:mass}(b) shows the wrong-sign invariant mass 
distribution of the $\lc K^+ \pi^-$ system  with the same binning as in (a).  
There is no significant excess. 
   
In Fig.~\ref{fig:mass}(c) the shaded region from (a) is re-plotted in 
2.5 $\MeV /c^2$ bins and fit with a maximum likelihood technique to a 
Gaussian plus linear background.  The fit has $\chi^2$/dof = 0.45, indicating 
that the background is linear in this region.

To determine the combinatoric background under the signal peak we exploit the 
linearity of the background justified by the fit.  We define symmetric regions 
of  the mass plot in Fig.~\ref{fig:mass}(c): 
(i) the signal region ($3520\pm5 \MeV /c^2$) with $\Nevt$ events; and 
(ii) 115 $\MeV /c^2$ sideband regions above and below the signal region, 
containing $\Ntot-\Nevt=\Nside$ events.  We estimate the number of expected 
background events in the signal region from the sidebands as 
$\Nside*5/(115)=\Nback \pm \dBack$ events.  This determination has a 
(Gaussian) statistical uncertainty, solely from counting statistics.
The single-bin significance of this signal is the excess in the signal region 
divided by the total uncertainty in the background estimate: 
$\Nsig / \surd(6.1 + 0.5^2) = 6.3\sigma$~\cite{SIG}.  The Poisson probability 
of observing at least this excess, including the Gaussian uncertainty in the 
background, is $1.0 \times 10^{-6}$.

Our reconstruction mass window is 3.2-4.3 $\GeV /c^2$ with 110 bins 
of width 10 $\MeV /c^2$ in this interval.  The overall probability of  
observing an excess at least as large as the one we see anywhere in the 
search interval is $1.1 \times 10^{-4}$.  

% Signal Properties
%

This state has a fit mass of $\Mass \pm \dMass$ $\MeV /c^2$. Our expected 
mass resolution, from a simulation of the decay $\ccdlcki$ is 
$\Res$ $\MeV /c^2$.  We observe a Gaussian width of $\Width\pm \dWidth$ 
$\MeV /c^2$, consistent with our simulation.  The confidence level for a fit 
with a Gaussian width fixed at our expected resolution is $20\%$.  The width 
we observe is consistent with statistical fluctuations in this small sample.

The wrong-sign mass combination is plotted in Fig.~\ref{fig:mass}(b).  
Those events show comparable fluctuations to the sidebands of the signal 
channel and give no evidence for a significant narrow structure.  We have 
investigated all possible permutations of mass assignments for the non-$\lc$ 
tracks.  The peak at 3520 $\MeV /c^2$ disappears for any other mass choice, 
and no other significant structures are observed.  Reconstructions with events 
from the $\lc$ mass sidebands produce relatively few  entries and no 
significant structures in the doubly charmed baryon region.

\begin{figure}[ht]
%\ref{fig:signf}
\center{\psfig{figure=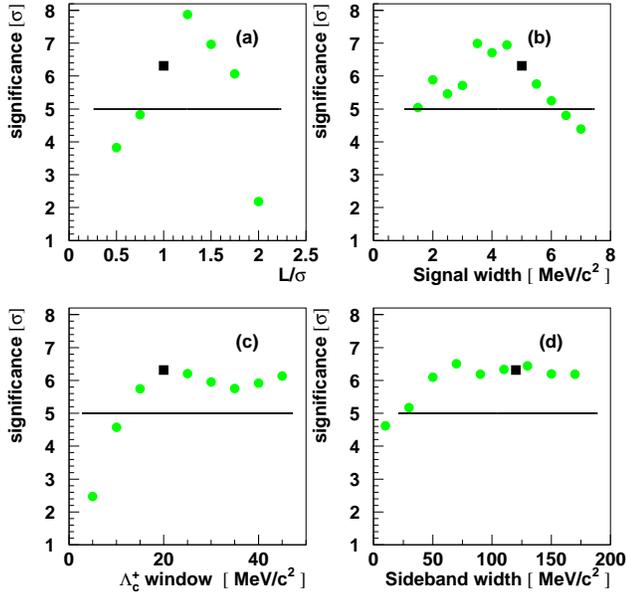,width=90mm}}
\caption{The signal significance as a function of: 
         (a) the vertex significance of the the $\kpi$ vertex, 
         (b) the half width of the signal region.
         (c) the mass window around the $\lc$ mass,
         (d) the half width of the sideband region.
         The square points are the values used in this analysis.}
\label{fig:signf}
\end{figure}

The dependence of the signal significance, as defined above, on several of the
selection cuts is shown in Fig.~\ref{fig:signf}.  The significance depends
strongly only on the $\kpi$ vertex separation.  The dependence is driven by a
large increase in background at small separations and the absence of both 
signal and background events at large separations.  The dependence on the width
of the signal region is stable, only decreasing when made wider than the 
mass resolution.  All other cuts have been checked; no significant dependence 
on any cut has been seen.

A weakly-decaying $\Xi_{cc}^+$ state has two $c$ quark decay amplitudes plus
a W-exchange amplitude for $c+d \rightarrow s+u$.  This suggests that
its lifetime will be of the order of the $\Xi_c^0$ or shorter, rather than
like the long-lived $\Xi_c^+$.  For each event we compute the reduced proper 
time \\
$t = (m/p)\times(L-L_{min})/c$.  Here $L$ is the flight 
path from the primary vertex to the $\Xi_{cc}^+$ decay point, and 
$L_{min}$ is the minimum flight path for a given event.  In this analysis 
$L_{min}$ = $\sigma$, the error on the flight path.  One can form the 
likelihood, $\mathcal{L(\tau)}$ = $\prod_{i=1}^{N_{ev}}$ P($t_i, \tau$)
to observe a set {$\{t_i\}$} of reduced proper time values, 
given the probability P($t, \tau$) for a given lifetime $\tau$. 
P($t, \tau$) is the normalized probability of observing a given reduced 
proper time $t$ for a true lifetime $\tau$ determined from  the 
SELEX simulation.

\begin{table}[!ht]
\centering
\begin{tabular}{|c|c|c|c|c|c|}
$\tau~(\fsec)$          &   0   &   12.5   &  25      &  37.5  & 50      \\
\hline
 signal region    &  0.2 &   0.0   &  0.58    &  1.54  & 2.70    \\
 sideband region events &  1.2 &   0.8   &  0.2    &  0.0  & 0.6    \\
\end{tabular}
\caption{The change in likelihood, 
$-log(\mathcal{L})$ + $log(\mathcal{L}$$_{max})$  for different 
$\Xi_{cc}^+$ lifetimes.  Increases of 0.5 and 2.0 in this statistic 
correspond to $1\sigma$ and $2\sigma$ 
respectively.}
\label{tab:life}
\end{table}

Results for the change in likelihood from its maximum value are given in 
Table~\ref{tab:life} for both the signal region and a sideband region, where 
combinatorial backgrounds dominate.  In both cases the lifetime is much 
shorter than the single-charm $\Xi_c^0$ lifetime of about 
100~$\fsec$~\cite{PDG}.  Given our single-event resolution of 
20~$\fsec$~\cite{E781-PRL}, we cannot exclude a prompt decay for the signal 
events from the likelihood.  However, the fact that the significance of the 
signal increases as the flight path L is increased [see Fig~\ref{fig:signf}(a)]
suggests that the lifetime is not zero.  From these data the upper limit for 
the $\Xi_{cc}^+$ lifetime at $90\%$ confidence is 33~$\fsec$.

\begin{table}[!ht]
\centering
\begin{tabular}{|l||c|c|c|c|}
                        & $\Sigma^{-}$ & $\pi^-$ & proton & $\pi^+$\\
\hline
 interaction fraction   &    0.67      &   0.13  &  0.18  &  0.01  \\
 signal region events   &    18        &   0     &  4     &  0     \\
 sideband region events &    110       &   7     &  21    &  2     \\
\end{tabular}
\caption{Events produced by incident particle species.}
\label{tab:beam-production}
\end{table}

\begin{table}[!ht]
\centering
\begin{tabular}{|l||c|c|c|c|}
                        &   Diamond    & Copper \\
\hline
 fraction of $\lc$ signal events      &    0.68      &  0.32  \\
 signal region events   &    18        &  4     \\
 sideband region events &    93        &  47    \\
\end{tabular}
\caption{Events produced by different target materials.}
\label{tab:target-production}
\end{table}

Production characteristics of the 22 signal plus background events are 
indistinguishable from the single-charm $\lc$ sample.~\cite{E781-PL}  The mean 
$p_t$ is 1 $\GeV /c$ and  mean $x_F \sim$ 0.33.  It is interesting to compare 
production of the $\Xi_{cc}^+$ state by different beam hadrons.  Results for 
the signal region and sidebands are listed in Table~\ref{tab:beam-production}. 
One sees that the doubly charmed baryon candidates are produced solely by the 
baryon beams.  The top row gives the fraction of all interactions produced 
from each of the beams.  The proton/$\Sigma^-$ ratio for $\Xi_{cc}^+$ 
production and for all interactions is the same.  One can also ask if there a 
dependence on the target nucleus.  Table~\ref{tab:target-production} shows 
that the diamond/copper ratio of the signal events is similar to the sideband 
events, which in turn behave like single-charm production.  Production of 
$\Xi_{cc}^+$ candidates differs from $\lc$ production only in the absence of 
meson beam contributions.

The yield of this state is larger than most production models 
predict~\cite{murray}.  The acceptance for the $\Nsig$ events we observe in 
this final state, given that we observe a $\lc$, is 11\%.  Using a factor 1.5 
from isospin to account for the $\ccdlcko$ mode and Bjorken's 
estimate~\cite{bj} of 1.6 to include other decay modes with $\lc$ in the 
final state, we find that $\sim$20\% of the $\lc$ in this sample are produced 
by $\ccd$ decay.  

Our production region has not been probed by other experiments.  The large 
$x_F$, small $p_t$ region is not amenable to perturbative QCD analysis.  
Two CERN experiments with $\pi^-$ beams have reported anomalously large 
production of events with two charmed particles~\cite{wa75,simon}.   However, 
we do not know of any model calculation that would predict this large 
hadro-production rate of doubly charmed events.  The FOCUS charm 
photo-production experiment at Fermilab has searched for doubly charmed 
baryons in their charm samples.  They see very few candidate events and no 
signal peaks~\cite{FOCUS}.  The BELLE B-factory experiment recently reported 
a very large production ratio for  (J/$\psi c\overline{c})$/(J/$\psi)$ in 
continuum $e^+$~$e^-$ annihilation at 10.8 $\GeV$~\cite{BELLE}.  The 
connection to the hadro-production data is not clear.

% Conclusions
%

In summary, we have observed a narrow state at 3520 $\MeV /c^2$ decaying 
into $\lc K^-\pi^+ $, consistent with the weak decay of the 
doubly charmed baryon $\Xi_{cc}^+$.  We report this state as the first 
observation of a doubly charmed baryon.

% Acknowledgement_paragraph.tex
%

The authors are indebted to the staff of Fermi National Accelerator Laboratory
and for invaluable technical support from the staffs of collaborating
institutions.
This project was supported in part by Bundesministerium f\"ur Bildung, 
Wissenschaft, Forschung und Technologie, Consejo Nacional de 
Ciencia y Tecnolog\'{\i}a {\nobreak (CONACyT)},
Conselho Nacional de Desenvolvimento Cient\'{\i}fico e Tecnol\'ogico,
Fondo de Apoyo a la Investigaci\'on (UASLP),
Funda\c{c}\~ao de Amparo \`a Pesquisa do Estado de S\~ao Paulo (FAPESP),
the Israel Science Foundation founded by the Israel Academy of Sciences and 
Humanities, Istituto Nazionale di Fisica Nucleare (INFN),
the International Science Foundation (ISF),
the National Science Foundation (Phy \#9602178),
NATO (grant CR6.941058-1360/94),
the Russian Academy of Science,
the Russian Ministry of Science and Technology,
the Turkish Scientific and Technological Research Board (T\"{U}B\.ITAK),
the U.S. Department of Energy (DOE grant DE-FG02-91ER40664 and DOE contract
number DE-AC02-76CHO3000), and
the U.S.-Israel Binational Science Foundation (BSF).


\begin{references}
\bibitem[\ast]{}deceased
\bibitem[\dag]{}Present address: Infinion, M\"unchen, Germany
\bibitem[\ddag]{}Now at Imperial College, London SW7 2BZ, U.K.
\bibitem[\S]{}Now at Instituto de F\'{\i}sica da Universidade Estadual de Campinas, UNICAMP, SP, Brazil
\bibitem[\P]{}Now at Physik-Department, Technische Universit\"at M\"unchen, 85748 Garching, Germany
\bibitem[\parallel]{}Present address: The Boston Consulting Group, M\"unchen, Germany
\bibitem[\ast\ast]{}Now at Instituto de F\'{\i}sica Te\'orica da Universidade Estadual Paulista, S\~ao Paulo, Brazil
\bibitem[\dag\dag]{}Present address: Lucent Technologies, Naperville, IL
\bibitem[\ddag\ddag]{}Present address: SPSS Inc., Chicago, IL
\bibitem[\S\S]{}Now at University of Alabama at Birmingham, Birmingham, AL 35294
\bibitem[\P\P]{}Present address: Imadent Ltd.,\ Rehovot 76702, Israel
\bibitem[\ast\ast\ast]{}Present address: DOE, Germantown, MD
\bibitem[\dag\dag\dag]{}Now at Solidum, Ottawa, Ontario, Canada
\bibitem[\ddag\ddag\ddag]{} Present address: Siemens Medizintechnik, Erlangen, Germany
\bibitem[\S\S\S]{}Present address: Deutsche Bank AG, Eschborn, Germany

\vskip 0.25cm

\bibitem{PDG}    Particle Data Group, D.E. Groom {\sl et al.}, 
                 Eur.\ Phys.\ J.\ {\bf C15,} 1 (2000).


\bibitem{HQEH}   A. DeRujula, H. Georgi, and S. Glashow,
                 Phys.\ Rev. {\bf D12}, 147 (1975).

\bibitem{SW}   M. Savage and M. Wise,
                 Phys.\ Lett. {\bf B248}, 177 (1990).

\bibitem{Korner} J.G. K\"{o}rner, M. Kr\"{a}mer, and D. Pirjol 
                 Prog.\ Part.\ Nucl. \ Phys.\ {\bf 33}, 787 (1994).

\bibitem{BPhys} See also references 111-124 in hep-ph/0201071,
B Physics at the Tevatron.
%%CITATION = PRPLC,289,1;%%


\bibitem{Thesis} M. Mattson, Ph.D. thesis, Carnegie Mellon University, 2002.

\bibitem{SIG} The statistical measure appropriate for a discovery claim is the 
              probability that a background fluctuation can account for all 
              events in the signal region.  This is different from  the yield 
              significance, which goes as $S/\sqrt(S+B)$, e.g. the area of a 
              Gaussian fit.  Applying a yield significance test to a discovery 
              claim would require $\ge25$ signal events in a background free 
              situation in order to claim a $5\sigma$ limit.

\bibitem{SELEX}  SELEX Collaboration, J.S. Russ {\sl et al.}, 
                 in {\it Proceedings of the 29th International Conference
                 on High Energy Physics}, 1998, 
                 edited by A. Astbury {\sl et al.} 
                 (World Scientific, Singapore, 1998), Vol. II, p. 1259;
                 hep-ex/9812031.

\bibitem{RICH}   J. Engelfried {\sl et al.}, 
                 Nucl.\ Instrum.\ Methods A {\bf 431}, 53 (1999).

\bibitem{E781-PRL} A.~Kushnirenko {\sl et al.}
                 Phys.\ Rev.\ Lett.\ {\bf 86}, 5243 (2001), hep-ex/0010014.

\bibitem{E781-PL} F.~Garcia {\sl et al.}
                 Phys.\ Lett.\ {\bf B528}, 49 (2002), hep-ex/0109017.

\bibitem{murray} M. ~Moinester, Zeit.\ Phys. {\bf A355},349 (1996), \\
                 V. Kiselev and A. Likhoded, hep-ex/0103169 (2001) 
                 and references therein.

\bibitem{bj} J. B. Bjorken, Fermilab-Conf-85/69.

\bibitem{wa75} S.~Aoki {\sl et al.}
                 Phys.\ Lett.\ {\bf B187}, 185 (1987).

\bibitem{simon} ACCMOR Collaboration, S. ~Barlag, {\sl et al.}
                 Phys.\ Lett.\ {\bf B257}, 519 (1991).

\bibitem{FOCUS} S. ~Ratti, BEACH2002, Vancouver, B.C., see also\\
www.hep.vanderbilt.edu/$\sim$stenson/xicc/xicc\_focus.html

\bibitem{BELLE} Belle Collaboration, K. Abe, {\sl et al.}, hep-ex/0205104

\end{references}
\end{document}